\documentclass[10pt,twocolumn,letterpaper]{article}

\usepackage{iccv}
\usepackage{times}
\usepackage{epsfig}
\usepackage{graphicx}
\usepackage{amsmath}
\usepackage{amssymb}
\usepackage{booktabs}
\usepackage{comment}
\usepackage{subfigure}
\usepackage[misc]{ifsym}
\usepackage[normalem]{ulem}
\usepackage{multirow}

\usepackage[pagebackref=true,breaklinks=true,letterpaper=true,colorlinks,bookmarks=false]{hyperref}

\iccvfinalcopy 

\def\iccvPaperID{****} 

\renewcommand\footnotemark{}

\begin{document}

\title{Remote Heart Rate Measurement from Highly Compressed Facial Videos: an End-to-end Deep Learning Solution with Video Enhancement}

\author{Zitong Yu\textsuperscript{1* \thanks{ * Equal contribution~~
    $^{\dag}$ Corresponding author~~}
    }, Wei Peng\textsuperscript{1*}, Xiaobai Li\textsuperscript{1}, Xiaopeng Hong\textsuperscript{2,1}, Guoying Zhao\textsuperscript{3,1\dag}\thanks{\textcolor{blue}{Copyright @ 2019 IEEE. Personal use of this material is
permitted. However, permission to use this material for any other purposes must be obtained from the IEEE.}}\\
\textsuperscript{1}Center for Machine Vision and Signal Analysis, University of Oulu, Finland\\
\textsuperscript{2}MOE Key Lab. for Intelligent Networks and Network Security\\Faculty of Electronic and Information Engineering, Xi'an Jiaotong University, PRC \\
\textsuperscript{3}School of Information and Technology
, Northwest University, PRC\\
{\tt\small \{zitong.yu, wei.peng, xiaobai.li, xiaopeng.hong, guoying.zhao\}@oulu.fi}
}


\maketitle

\begin{abstract}

Remote photoplethysmography (rPPG), which aims at measuring heart activities without any contact, has great potential in many applications (e.g., remote healthcare). Existing rPPG approaches rely on analyzing very fine details of facial videos, which are prone to be affected by video compression. Here we propose a two-stage, end-to-end method using hidden rPPG information enhancement and attention networks, which is the first attempt to counter video compression loss and recover rPPG signals from highly compressed videos. The method includes two parts: 1) a Spatio-Temporal Video Enhancement Network (STVEN) for video enhancement, and 2) an rPPG network (rPPGNet) for rPPG signal recovery.  The rPPGNet can work on its own for robust rPPG measurement, and the STVEN network can be added and jointly trained to further boost the performance especially on highly compressed videos. Comprehensive experiments are performed on two benchmark datasets to show that, 1) the proposed method not only achieves superior performance on compressed videos with high-quality videos pair, 2) it also generalizes well on novel data with only compressed videos available, which implies the promising potential for real-world applications.

\end{abstract}

\vspace{-0.6em}
\section{Introduction}

\begin{figure}
\includegraphics[width=8cm,height=6.2cm]{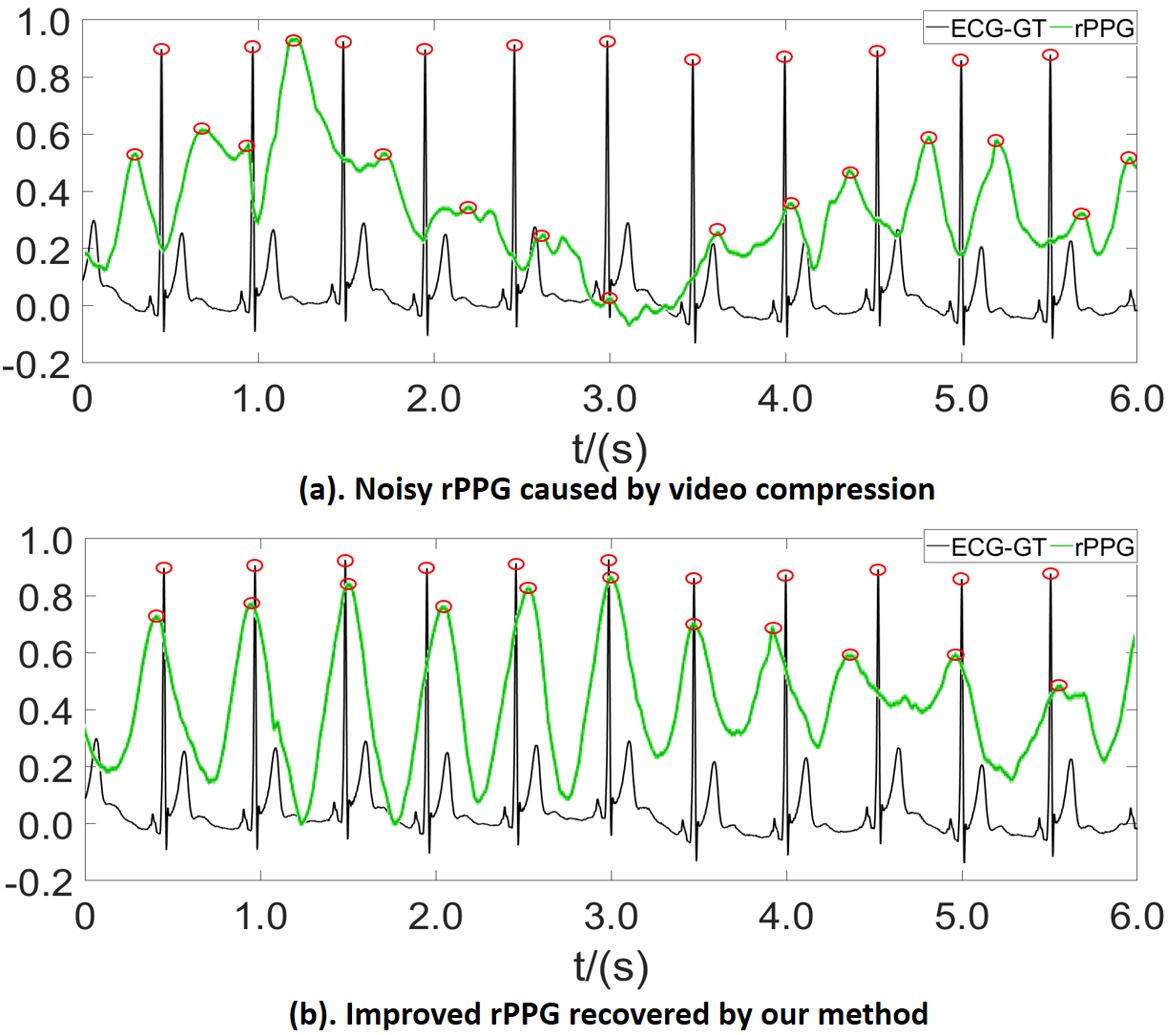}
  \caption{\small{
  rPPG measurement from highly compressed videos. Due to video compression artifact and rPPG information loss, the rPPG in (a) has very noisy shape and inaccurate peak counts which lead to erroneous heart rate measures, while after video enhancement by STVEN, the rPPG in (b) shows more regular pulse shape with accurate peak locations comparing to the ground truth ECG.}
  }
   
\label{fig:rppg_mainidea}
\vspace{-0.6em}
\end{figure}

Electrocardiography (ECG) and Photoplethysmograph (PPG) provide common ways for measuring heart activities. These two types signals are important for healthcare applications since they provide the measurement of both basic average heart rate (HR) and more detailed information like heart rate variability (HRV). However, these signals are mostly measured from skin-contact ECG/BVP sensors, which may cause discomfort and are inconvenient for long-term monitoring. To solve this problem, remote photoplethysmography (rPPG) , which targets to measure heart activity remotely and without any contact, has been developing rapidly in recent years \cite{CHROM,Li2014,Poh2010,Poh2011,Tulyakov2016,Verkruysse:08,gang2019Atrial}. 

However, most previous rPPG measurement works did not take the influence of video compression into consideration, whereas the fact is that most videos captured by commercial cameras are compressed through different compression codecs with various bitrates. 
Recently, two works~\cite{hanfland2016video,mcduff2017impact} pointed out and demonstrated that the performance of rPPG measurement dropped to various extents when using compressed videos with different bitrates.
As shown in Fig. \ref{fig:rppg_mainidea}(a), rPPG signals measured from highly compressed videos usually suffer from noisy curve shape and inaccurate peak locations due to information loss caused by both intra-frame and inter-frame coding of the video compression process. 
Video compression is inevitable for remote services considering the convenient storage and transmission in Internet. Thus it is of great practical value to develop rPPG methods that can work robustly on highly compressed videos. However, no solution has been proposed yet to counter this problem.


To address this problem, we propose a two-stage, end-to-end method using hidden rPPG information enhancement and attention networks, which can counter video compression loss and recover rPPG signals from highly compressed facial videos. Figure \ref{fig:rppg_mainidea}(b) illustrates the advantages of our method on rPPG measurement from highly compressed videos. Our contributions include:


\begin{itemize}
\setlength\itemsep{-0.1em}
\vspace{-0.5em}
    \item To our best knowledge, we provide the first solution for robust rPPG measurement directly from compressed videos, which is an end-to-end framework made up of a video enhancement module STVEN (Spatio-Temporal Video Enhancement Network) and a powerful signal recovery module rPPGNet.

    \item The rPPGNet, featured with a skin-based attention module and partition constraints, can measure accurately at both HR and HRV levels. Compared with previous works which only output simple HR numbers\cite{Xuesong2018, HR-CNN}, the proposed rPPGNet produces much richer rPPG signals with curve shapes and peak locations. Moreover, It outperforms state-of-art methods on various video formats of a benchmark dataset even without using the STVEN module.

    \item The STVEN, which is a video-to-video translation generator aided with fine-grained learning, is the first video compression enhancement network to boost rPPG measurement on highly compressed videos. 

    \item We conduct cross-dataset test and show that the STVEN can generalize well to enhance unseen, highly compressed facial videos for robust rPPG measurement, which implies promising potential in real-world applications.
    
\end{itemize}

\noindent
\vspace{-0.3em}

    

    
\vspace{-1.5em}
\section{Related Work}


\noindent\textbf{Remote Photoplethysmography Measurement.}\quad      In past few years, several traditional methods explored rPPG measurement from videos by analyzing subtle color changes on facial regions of interest (ROI), including blind source separation~\cite{Poh2010, Poh2011}, least mean square~\cite{Li2014}, majority voting~\cite{lam2015robust} and self-adaptive matrix completion~\cite{Tulyakov2016}. However, ROI selection in these works were customized or arbitrary, which may cause information loss. Theoretically speaking, all skin pixels can contribute to the rPPG signals recovery. There are other traditional methods which utilized all skin pixels for rPPG measurement, e.g., chrominance-based rPPG (CHROM)~\cite{CHROM}, projection plane orthogonal to the skin tone (POS)~\cite{wang2017algorithmic}, and spatial subspace rotation~\cite{2SR, wang2010locality}. All these methods treat each skin pixel with equal contribution, which is against the fact that different skin parts may bear different weights for rPPG singals recovery.





More recently, a few deep learning based methods were proposed for average HR estimation, including SynRhythm~\cite{Xuesong2018}, HR-CNN~\cite{HR-CNN} and DeepPhys~\cite{DeepPhys}. Convolutional neural networks (CNN) were also employed for skin segmentation~\cite{chaichulee2017multi,tang2018non} and then to predict HR from skin regions. These methods were based on spatial 2D CNN, which failed to capture temporal features which are essential for rPPG measurement. Moreover, the skin segmentation task was treated separately from the rPPG recovery task, which lacks the mutual feature sharing between such two highly related tasks.



\noindent\textbf{Video Compression and Its Impact for rPPG.}\quad   In real-world applications, video compression is widely used because of its great storage capacities with minimal quality degradation. Numerous codecs for video compression have been developed as standards of the Moving Picture Experts Group (MPEG) and International Telecommunication Union Telecommunication Standardization Sector (ITU-T). These include MPEG-2 Part 2/H.262 ~\cite{H262} and the low bitrate standard MPEG-4 Part 2/H.263~\cite{puri1998mpeg}. Current-generation standard AVC/H.264 ~\cite{wiegand2003overview} achieves an approximate doubling in encoding efficiency over H.262 and H.263. More recently, next-generation standard HEVC/H.265 ~\cite{sullivan2012overview} utilizes increasingly complex encoding strategies for an approximate doubling in encoding efficiency over H.264. 

In the stage of video coding, compression artifacts are inevitable as a result of quantization. Specifically, the existing compression standards drop subtle changes that human eyes cannot see. It does not favor the purpose of rPPG measurement, which mainly relies on subtle changes at invisible level.
The impact of video compression on rPPG measurement was not explored until very recently. Three works\cite{hanfland2016video,mcduff2017impact,spetlik2018non} consistently demonstrated that the compression artifacts do reduce the accuracy of HR estimation. However, these works only tested on small-scale private datasets using traditional methods, and it was unclear whether compression also impacted deep learning based rPPG methods on large dataset. Furthermore, these works just pointed out the problem of compression on rPPG, but no solution has been proposed yet.




\begin{figure*}[t]
\centering
\includegraphics[width=0.97\textwidth]{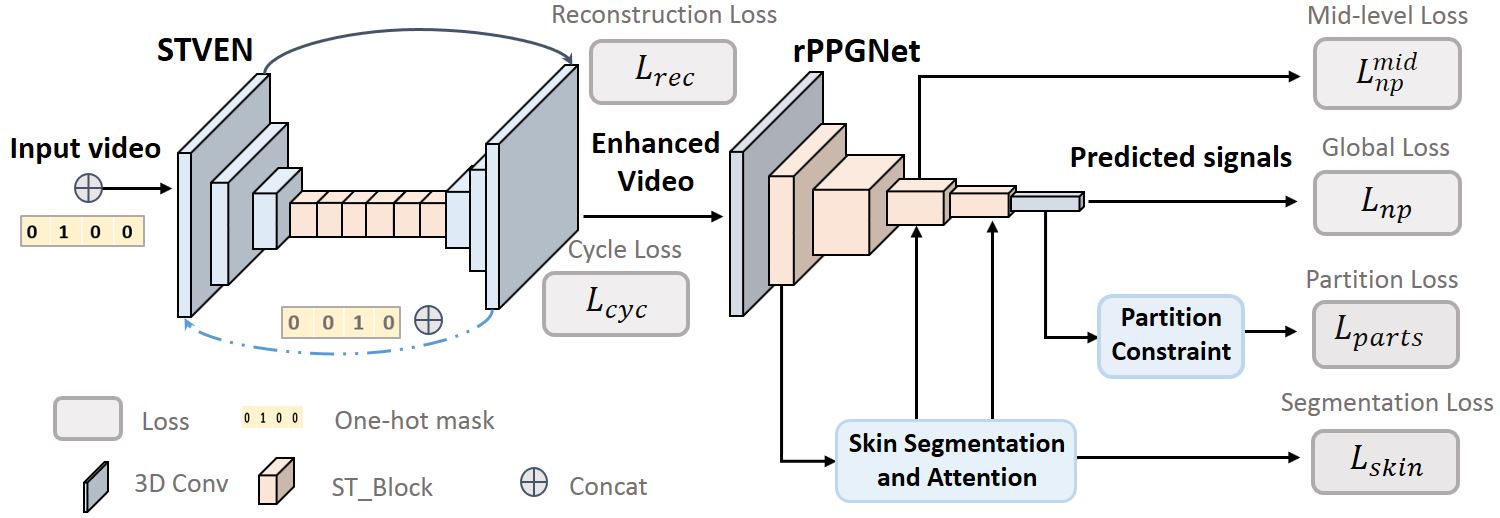}

\caption{\small{Illustration of the overall framework. There are two models in our framework: video quality enhancement model STVEN (left) and rPPG recovery model rPPGNet (right). Both of them work well by learning with corresponding loss functions. We will also introduce an elaborate joint training, which further improves the rPPG recovery performance.}}

\label{fig:framework}
\vspace{-1em}
\end{figure*}


\noindent\textbf{Quality Enhancement for Compressed Video.}\quad           Fueled by the high performance of deep learning, several works introduce it to enhance the quality of compressed videos and get promising results, including ARCNN~\cite{dong2015compression}, deep residual denoising  neural networks (DnCNN)~\cite{zhang2017beyond}, generative adversarial networks~\cite{galteri2017deep} and multi-frame quality enhancement network~\cite{yang2018multi}. 
However, all of them were designed for solving general compression problems or other tasks like object detection, but not for rPPG measurement. 
There are two works~\cite{mcduff2018deep,zhao2018novel} about rPPG recovery from low quality videos. The ~\cite{mcduff2018deep} focused on frame resolutions but not about video compression and format. The other one~\cite{zhao2018novel} tried to address the rPPG issue on compressed videos, but the approach was only on bio-signal processing level AFTER the rPPG was extracted, which has nothing to do with video enhancement. To the best of our knowledge, no video enhancement method has ever been proposed for the problem of rPPG recovery from highly compressed videos. 


In order to overcome the above-mentioned drawbacks and fill in the blank, we propose a two-stage, end-to-end deep learning based method for rPPG measurement from highly compressed videos.



\section{Methodology}
\label{sec:method}

As a two-stage end-to-end method, we will first introduce our video enhancement network STVEN in Section~\ref{sec:STVEN}, then introduce the rPPG signal recovery network rPPGNet in Section~\ref{sec:rPPGNet}, and at last explain how to jointly train these two parts for boosting performance. The overall framework is shown in Fig~\ref{fig:framework}.


\subsection{STVEN}
\label{sec:STVEN}


For the sake of enhancing the quality of highly compressed videos, we present a video-to-video generator called Spatio-Temporal Video Enhancement Networks (STVEN), which is shown in the left of Fig.\ref{fig:framework}. Here we perform a fine-grained learning by assuming that compression artifacts from different compression bitrates are with different distributions. As a result, compressed videos are placed into the buckets $[0,1,2,...,\mathcal{C}]$ denoted as $\mathbb{C}$ based on their compression bitrate. Here, 0 and $\mathcal{C}$ represent videos with lowest and highest compression rate, respectively. Let $c_{k}^{\tau} = [c_{k1},c_{k2},...,c_{k\tau}]$ be a sequence of the compressed video with length of $\tau$ for $k \in \mathbb{C}$. Then our goal is to train a generator $\mathcal{G}$ which can enhance the quality of compressed videos $c_{k}^{\tau}$ so that the distribution of the video is identical to the one of which $k=0$, that is original video $c_{0}^{\tau}$. Let say the output of generator $\mathcal{G}$ is $\hat{c}_{0}^{\tau} = [\hat{c}_{01},\hat{c}_{02},...,\hat{c}_{0\tau}]$. Then the conditional distribution of $\hat{c}_{0}^{\tau}$ given input videos $c_{k}^{\tau}$ and  video quality target 0 should be equal to the ${c}_{0}^{\tau}$ given input videos $c_{k}^{\tau}$ and target 0. That is
\begin{equation}\label{eq:contional}
p(\hat{c}_{0}^{\tau}|c_{k}^{\tau},0)  = p({c}_{0}^{\tau}|c_{k}^{\tau},0). 
\end{equation}
By learning to match the video distributions, our model generates the video sequences with the quality being enhanced. Likewise, in order to make the model more generalizable, the framework is also set to be able to compress the original video with a specific compression bitrate. This means that when our model is fed with video ${c}_{0}^{\tau}$ and outputs lower quality target $k$, the model $\mathcal{G}$ should also be able to generate the video which fits the distribution with the specific compression bitrate $k$. That is
\begin{equation}\label{eq:contional}
 p(\hat{c}_{k}^{\tau}|{c}_{0}^{\tau},k) = p({c}_{k}^{\tau}|{c}_{0}^{\tau},k),
\end{equation}
here $\hat{c}_{k}^{\tau}$ is the output of our generator with the inputs ${c}_{0}^{\tau}$ and $k$. Therefore, there will be two parts of the loss function $L_{rec}$ in STVEN: one is the translation reconstruction loss, for which we introduce a mean squared error (MSE) to deal with the lost video details, and the other one is the lose for compression reconstruction, here we employ a L1 loss for it. Then
\begin{equation}\label{eq:loss_recon}
\begin{split}
L_{rec} = 
&{E}_{k \sim \mathbb{C},t}({c}_{0}^{\tau}(t)-\mathcal{G}(c_{k}^{\tau},0)(t))^{2}  \\ 
&+ {E}_{k \sim \mathbb{C},t}||{c}_{k}^{\tau}(t)- \mathcal{G}(c_{0}^{\tau},k)(t)||\\
\end{split}
\end{equation}
Here $t\in [1,\tau]$  is the $t$-th frame of the output video. In addition, like in~\cite{CycleGAN2017}, we also introduce a cycle-loss for better reconstruction. In this way, we expect our model to satisfy this case: when taking  ($\hat{c}_{0}^{\tau}$) of $\mathcal{G}$, which is fed with ${c}_{k}^{\tau}$ and the specific compression bitrate label $0$, and the compression bitrate label $k$ as its inputs, the following output should match the distribution of the initial input videos. Similarly, we perform the cycle processing for original video. As a result, the cycle loss $L_{cyc}$ in STVEN is
\begin{equation}\label{eq:loss_cyc}
\begin{split}
L_{cyc} = &{E}_{k \sim \mathbb{C},t}||{c}_{k}^{\tau}(t)- \mathcal{G}(\mathcal{G}(c_{k}^{\tau},0),k)(t)||\\
&+ {E}_{k \sim \mathbb{C},t}||{c}_{0}^{\tau}(t)- \mathcal{G}(\mathcal{G}(c_{0}^{\tau},k),0)(t)||.
\end{split}
\end{equation}
 Therefore, the total loss of STVEN $L_{STVEN}$ is the sum of $L_{rec}$ and $L_{cyc}$. To achieve this goal, we build our model STVEN with a spatial-temporal convolutional neural network. The architecture is composed of two downsampling layers and two upsampling layers at the two ends, with six spatio-temporal blocks in the middle. The details of the architecture is shown in the top of Table.~\ref{tab:network}.


\begin{table}
\begin{center}
\scalebox{0.85}{
\begin{tabular}{llcc}
\hline
& Layer & Output size & Kernel size \\
\hline\hline
\multirow{7}{*}{\textbf{STVEN}}
&Conv\_1 & $64\times T\times 128\times 128$ & $3\times 7\times 7$\\
&Conv\_2 & $128\times T\times 64\times 64$ & $3\times 4\times4$ \\
&Conv\_3 & $512\times \frac{T}{2} \times 32\times 32$ & $4\times 4\times 4$ \\
&ST\_Block & $512\times \frac{T}{2} \times 32\times 32$ & $\begin{bmatrix}
3\times 3\times 3
\end{bmatrix} \times 6$ \\
&DConv\_1 & $128\times T \times 64\times 64$ & $4\times 4\times 4$ \\
&DConv\_2 & $64\times T\times 128\times 128$ & $1\times 4\times4$ \\
&DConv\_3 & $3\times T\times 128\times 128$ & $1\times 7\times 7$\\
\hline\hline
\multirow{4}{*}{\textbf{rPPGNet}}
&Conv\_1 & $32\times T\times 64\times 64$ & $1\times 5\times 5$\\
&ST\_Block & $64\times T \times 16\times 16$ & $\begin{bmatrix}
3\times 3\times 3
\end{bmatrix} \times 4$ \\
&SGAP & $64\times T\times 1\times 1$ & $1\times 16\times 16$\\
&Conv\_2 & $1\times T\times 1\times 1$ & $1\times 1\times 1$\\
\hline
\hline
\end{tabular}}
\end{center}
\caption{\small{{The architecture of STVEN and rPPGNet.}   Here "Conv\_x" means 3D convolution filters and "DConv\_x" denotes 3D transposed convolution filters. "ST\_Block" represents spatio-temporal block~\cite{tran2018closer}, which is constructed by two sets of cascaded 3D convolution filters with kernel size of $1\times 3\times 3$ and $3\times 1\times 1$, respectively. Besides, we introduce instance normalization and ReLU into STVEN while batch normalization and ReLU into rPPGNet. "SGAP" is short for spatial global average pooling.}}
\label{tab:network}
\end{table}

\subsection{rPPGNet}
\label{sec:rPPGNet}
The proposed rPPGNet is composed of a  spatio-temporal convolutional network, a skin-based attention module and a partition constraint module. Skin-based attention helps to adaptively selected skin regions, and partition constraint is introduced for learning better rPPG feature representation. 

\textbf{Spatio-Temporal Convolutional Network.}\quad  Previous works like~\cite{CHROM, wang2017algorithmic},  usually projected spatial pooled RGB into another color space for better representation of the rPPG information. Then temporal context based normalization was used to get rid of irrelevant info (e.g., noise caused by illumination or motion). Here we merge these two steps into one model and propose an end-to-end spatio-temporal convolutional network, which takes \(T\)-frame face images with RGB channels as the inputs and outputs rPPG signals directly. The backbone and architecture of rPPGNet is shown in Fig.~\ref{fig:framework} and Table.~\ref{tab:network} respectively.



Aiming to recover rPPG signals $y\in \mathbb{R}^{T}$, which should have accurate pulse peak locations compared with the corresponding ground truth ECG signals $y^{g}\in \mathbb{R}^{T}$, negative Pearson correlation is used to define the loss function. It can be formulated as

\vspace{-1.5em}
\begin{equation} \small
L_{np}=1-\frac{T\!\sum\limits_{i=1}^{T}y_{i}y_{i}^{g}-\sum\limits_{i=1}^{T}y_{i}\sum\limits_{i=1}^{T}y_{i}^{g}}{\sqrt{(T\!\sum\limits_{i=1}^{T}y_{i}^2\!-\!(\sum\limits_{i=1}^{T}y_{i})^2)(T\!\sum\limits_{i=1}^{T}(y_{i}^{g})^2\!-\!(\sum\limits_{i=1}^{T}y_{i}^{g})^2)}}.
\vspace{-0.5em}
\end{equation}
Unlike Mean Square Error (MSE), our loss is to minimize the linear similarity error instead of the point-wise intensity error. We tried MSE loss in prior test, which achieved much worse performance because the intensity values of signals are irrelevant with our task (i.e., to measure accurate peak locations) and introduces extra noise inevitably.

We also aggregate the mid-level features (outputs of the third ST\_Block) into pseudo signals and then constrain them by $L_{np}^{mid}$ for stable convergence. So the basic learning object for recovering rPPG singals is described as

\vspace{-0.5em}
\begin{equation} 
L_{rPPG}=\alpha L_{np}+\beta L_{np}^{mid},
\end{equation}
where $\alpha$ and $\beta$ are the weights for balancing the loss.

\begin{figure}
\centering
\includegraphics[width=8.2cm,height=3.8cm]{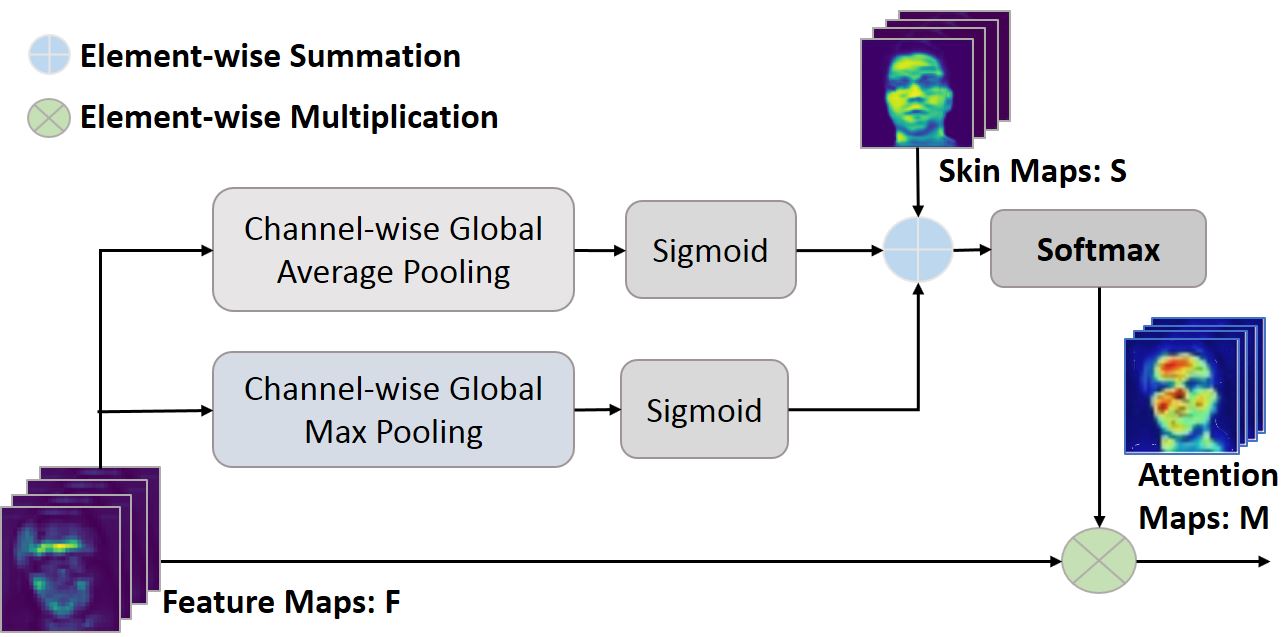}
\caption{Illustration of the skin-based attention module of the rPPGNet, which is parameter-free. It assigns importance to different locations in accordance with both skin confidence and rPPG feature maps. The softmax operation can be either spatial-wise or spatio-temporal-wise.}
\label{fig:attention}
\vspace{-1em}
\end{figure}

\textbf{Skin Segmentation and Attention.}\quad
Various skin regions have varying density degrees of blood vessels as well as biophysical parameter maps (melanin and haemoglobin), thus contribute at different levels for rPPG signal measurement. So the skin segmentation task is highly related to rPPG signals recovery task. These two tasks can be treated as a multi-task learning problem. Thus we employ a skin segmentation branch after the first ST\_Block. The skin segmentation branch projects the shared low-level spatio-temporal features into skin domain, which is implemented by spatial and channel-wise convolutions with residual connections. As there is no ground truth skin map in related rPPG datasets, we generate the binary labels for each frame by adaptive skin segmentation algorithms~\cite{taylor2014adaptive}. With these binary skin labels, the skin segmentation branch is able to predict high quality skin maps $S\in \mathbb{R}^{T\times H\times W}$. Here we adopt binary cross entropy $L_{skin}$ as the loss function.



In order to eliminate the influence of non-skin regions and enhance dominant rPPG features, we construct a skin-based parameter-free attention module which refines the rPPG features by predicted attention maps $M\in \mathbb{R}^{T\times H\times W}$. The module is illustrated in Fig.~\ref{fig:attention} and the attention maps are computed as       
\vspace{-0.1em}
\begin{equation}  \small
M(F,S)=\varsigma (\sigma (AvgPool(F))+\sigma (MaxPool(F))+S),
\vspace{-0.1em}
\end{equation}
where $S$ and $F$ donote the predicted skin maps and rPPG feature maps respectively. $\sigma$ and $\varsigma$ represent the sigmoid and softmax function respectively.

\begin{figure}
\centering
\includegraphics[width=7cm,height=3.1cm]{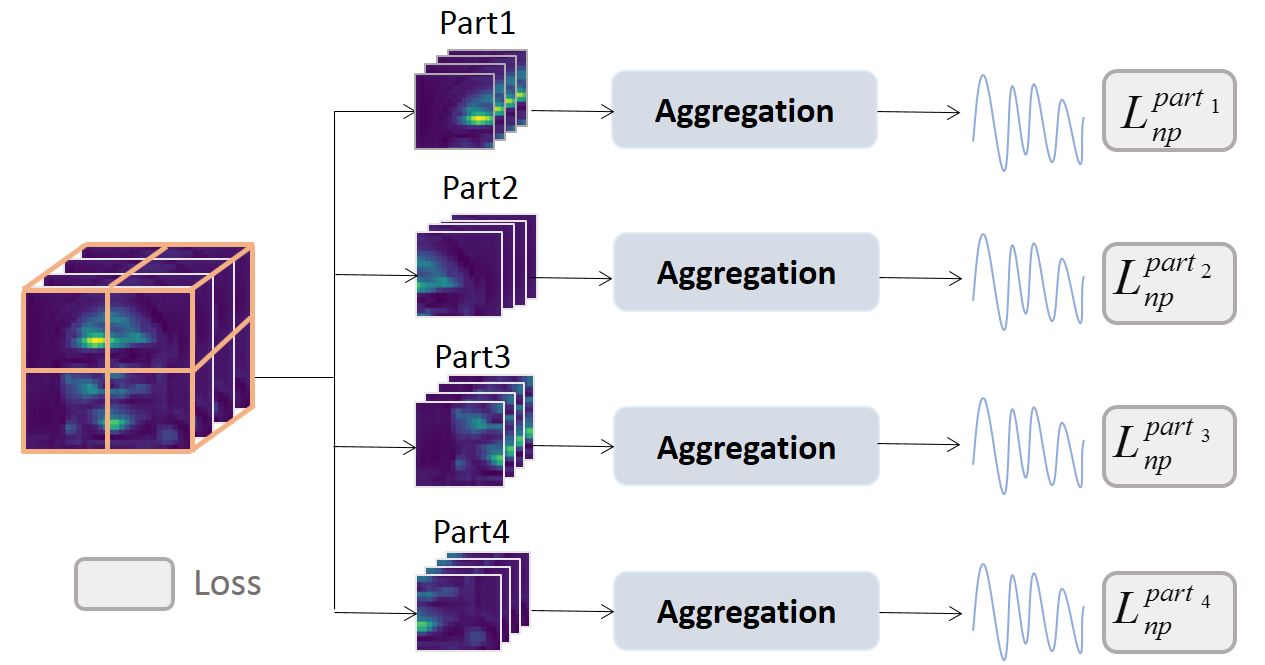}
\caption{Partition constraints with $N=4$.}
\label{fig:partition}
\vspace{-1em}
\end{figure}



\vspace{0.2em}
\textbf{Partition Constraint.}\quad       In order to help the model learn more concentrated rPPG features, local partition constraint is introduced. As shown in Fig.~\ref{fig:partition}, the deep features $D\in \mathbb{R}^{C\times T\times H\times W}$ are divided into $N$ uniform spatio-temporal parts $D_{i}\in \mathbb{R}^{C\times T\times (H/\sqrt{N})\times (W/\sqrt{N})}$, $i\in \left \{ 1,2,...,N \right \}$. Afterwards, spatial global average pooling is adopted by each part-level feature for feature aggregation and an independent $1\times1\times1$ convolution filter is deployed for final signals prediction. The partition loss is described as $L_{parts}=\sum_{i=1}^{N} L_{np}^{part_{i}}$, where $L_{np}^{part_{i}}$ is the negative Pearson loss of the $i$-th part-level feature.

The partition loss can be considered as a dropout ~\cite{srivastava2014dropout} for high-level features. It has a regularization effect because each partition loss is independent to each other, thus forcing part features to be powerful enough to recover the rPPG signal. In other words, via the partition constraint, the model can focus more on the rPPG signals instead of interference.

In sum, the loss function of rPPGNet can be written as
\begin{equation}  \small
\label{eq:rppg}
L_{rPPGNet}=L_{rPPG}+\gamma L_{skin}+\delta L_{parts},
\vspace{-0.2em}
\end{equation}
where $\gamma$ and  $\delta$ are the weights for balancing the loss.


\subsection{Joint Loss Training}


When STVEN is trained separately from rPPGNet, the output video cannot guarantee its effectiveness for the latter. Inspired by ~\cite{liu2017image}, we design an advanced joint training strategy to ensure that STVEN can enhance the video specifically in favor of rPPG recovery, which boosts the performance of rPPGNet even on highly compressed video. 



{First, we train the rPPGNet on the high quality videos with the training method described in Section~\ref{sec:rPPGNet}. Second, we train the STVEN on compressed videos with different bitrates. Finally, we train the cascaded networks, which is illustrated in Fig.~\ref{fig:framework}, with all high-level task model parameters fixed. Therefore, all the following loss functions are designed for the updating of STVEN.  Here we employ an application-oriented joint training, where we prefer the end-to-end performance rather than the performance of both stages. In this training strategy, we take away the cycle-loss part since we expect STVEN to recover richer rPPG signals instead of irrelevant information loss during video compression. As a result, we only need to know its target label, and the compression labels of all input videos fed into STVEN can be simply set to 0 as default. This allows the model to be more generalizable since it does not require subjectively compression labeling of input videos, thus can work on novel videos with unclear compression rate. Besides, like~\cite{johnson2016perceptual}, we also introduce a perceptual loss $L_p$ for joint training. That is}
\begin{equation}\label{eq:percep}
\vspace{-0.2em}
\begin{split}
L_p = 
&\frac{1}{T_{f}W_{f}H_{f}}\sum_{t=1}^{T_{f}}\sum_{i=1}^{W_{f}}\sum_{j=1}^{H_{f}}(\phi({c}_{0}^{\tau})(t,i,j)\\
&-\phi(\mathcal{G}(c_{k}^{T},0))(t, i,j))^{2}. 
\end{split}
\vspace{-0.2em}
\end{equation}

\begin{table*}[t]\footnotesize
\begin{center}
\caption{Performance comparison on OBF. HR is the averaged heart rate within 30 seconds, RF, LF, HF and LF/HF are HRV features that require finer inter-beat-interval measurement of rPPG signals. Smaller RMSE and bigger R values indicate better performance.  "rPPGNet\_base" denotes the spatio-temporal networks with $L_{rPPG}$ constraint, while "Skin", "Parts" and "Atten" indicate corresponding modules of rPPGNet described in Section~\ref{sec:rPPGNet}. "rPPGNet (full)" includes all modules of the rPPGNet.}


\label{tab:OBFrPPGNet}
\begin{tabular}{p{3.1cm} p{0.5cm} p{0.5cm} p{0.6cm} p{0.5cm} p{0.5cm} p{0.6cm}  p{0.5cm} p{0.5cm} p{0.6cm} p{0.5cm} p{0.5cm} p{0.6cm} p{0.5cm} p{0.5cm} p{0.6cm}}

\toprule
& \multicolumn{3}{c}{HR(bpm)} &  \multicolumn{3}{c}{RF(Hz)} &  \multicolumn{3}{c}{LF(u.n)} &  \multicolumn{3}{c}{HF(u.n)} &  \multicolumn{3}{c}{LF/HF}\\
  
\cmidrule(lr){2-4} \cmidrule(lr){5-7}
\cmidrule(lr){8-10} \cmidrule(lr){11-13}
\cmidrule(lr){14-16}

  \multicolumn{1}{c}{Method} & \multicolumn{1}{c}{SD} & RMSE & \multicolumn{1}{c}{R}  &  \multicolumn{1}{c}{SD} & RMSE & \multicolumn{1}{c}{R}  &  \multicolumn{1}{c}{SD} & RMSE & \multicolumn{1}{c}{R}  & \multicolumn{1}{c}{SD} & RMSE & \multicolumn{1}{c}{R}  &  \multicolumn{1}{c}{SD} & RMSE & \multicolumn{1}{c}{R}\\

 \midrule
 ROI\_green~\cite{li2018obf}   & 2.159 & 2.162 & 0.99   & 0.078 & 0.084  & 0.321  & 0.22  & 0.24 & 0.573 & 0.22  & 0.24 & 0.573 & 0.819  & 0.832 & 0.571\\
 
 CHROM~\cite{CHROM}   & 2.73 & 2.733 & 0.98   & 0.081 & 0.081  & 0.224  & 0.199  & 0.206 & 0.524 & 0.199  & 0.206 & 0.524 & 0.83  & 0.863 & 0.459\\
 
 POS~\cite{wang2017algorithmic}   & 1.899 & 1.906 & 0.991   & 0.07 & 0.07  & 0.44  & 0.155  & 0.158 & 0.727 & 0.155  & 0.158 & 0.727 & 0.663  & 0.679 & 0.687\\
 
 \midrule
  rPPGNet\_base  & 2.729 & 2.772 & 0.98  & 0.067 & 0.067  & 0.486 & 0.151  & 0.153 & 0.748 & 0.151  & 0.153 & 0.748 & 0.641  & 0.649 & 0.724\\
  
  rPPGNet\_base+Skin  & 2.548 & 2.587 & 0.983  & 0.067 & 0.067  & 0.483 & 0.145  & 0.147 & 0.768 & 0.145  & 0.147 & 0.768 & 0.616  & 0.622 & 0.749\\
  
  rPPGNet\_base+Skin+Parts  & 2.049 & 2.087 & 0.989  & 0.065 & 0.065  & 0.505 & 0.143  & 0.144 & 0.776 & 0.143  & 0.144 & 0.776 & 0.594  & 0.604 & 0.759\\

  rPPGNet\_base+Skin+Atten  & 2.004 & 2.051 & 0.989  & 0.065 & 0.065  & 0.515 & 0.137  & 0.139 & 0.79 & 0.137  & 0.139 & 0.79 & 0.591  & 0.601 & 0.76\\

 rPPGNet (full)  & \textbf{1.756} & \textbf{1.8} & \textbf{0.992}  & \textbf{0.064} & \textbf{0.064}  & \textbf{0.53} & \textbf{0.133}  & \textbf{0.135} & \textbf{0.804} & \textbf{0.133}  & \textbf{0.135} & \textbf{0.804} & \textbf{0.58}  & \textbf{0.589} & \textbf{0.773}\\
\bottomrule
\end{tabular}
\end{center}
\vspace{-2em}

\end{table*}
Here, $\phi$ denotes a differentiable function in rPPGNet and the feature maps $\phi(x) \in \mathbb{R}^{T_{f}\times W_{f}\times H_{f}}$. 
Cost function in Eq.~(\ref{eq:percep}) keeps the recovered video and the original video consistent in the feature map space. 
Besides, we also let STVEN contribute directly to rPPG task by introducing $L_{rPPG}$ as in Eq.~(\ref{eq:rppg}). In the joint training, we use the rPPG signals recovered from high quality videos as a softer target for the updating of STVEN, and it converges faster and more steadily than using the ECG signals, which might be too far-fetched and challenging as the target for highly compressed videos, as our prior tests proved. In all, the joint cost function $L_{joint}$ for STVEN can be formulated as
\vspace{-0.2em}
\begin{equation}\label{eq:joint}
L_{joint} = L_{rPPGNet} + \varepsilon L_{p} + \rho  L_{STVEN},
\end{equation}
\vspace{-0.2em}
here $\varepsilon$ and $\rho $ are hyper-parameters.

\vspace{-0.1em}
\section{Experiments}
\vspace{-0.1em}
\label{sec:experiemnts}
We test the proposed system in four sub-experiments, the first three on OBF ~\cite{li2018obf} dataset and the last one on MAHNOB-HCI~\cite{soleymani2012multimodal} dataset. Firstly, we evaluate the rPPGNet on OBF for both average HR and HRV feature measurement. Secondly, we compress OBF videos and explore how video compression influence the rPPG measurement performance. Thirdly, we demonstrate that STVEN can enhance the compressed videos and boost the rPPG measurement performance on OBF. Finally, we cross test the joint system of STVEN and rPPGNet on  MAHNOB-HCI, which has only compressed videos, to validate the generalizability of the system.

\vspace{-0.1em}
\subsection{Datasets and Settings}
\vspace{-0.1em}
\label{sec:dataset}
Two datasets - \textbf{OBF} ~\cite{li2018obf} and \textbf{MAHNOB-HCI} ~\cite{soleymani2012multimodal} are used in our experiments. The OBF is a recently release dataset for study about remote physiological signal measurement. It contains 200 five-minute-long RGB videos recorded from 100 healthy adults and the corresponding ground truth ECG signals are also provided. The videos are recorded at 60 fps with resolution of 1920x2080, and compressed in MPEG-4 with average bitrate $\approx$ 20000 kb/s (file size $\approx$ 728 MB). The long videos are cut into 30-seconds-long clips for our training and testing. The MAHNOB-HCI dataset is one of the most widely used benchmark for remote HR measurement evaluations. It includes 527 facial videos with corresponding physiological signals from 27 subjects. The videos are recorded with 61 fps with resolution of 780x580, which are compressed in AVC/H.264, average bitrate $\approx$ 4200 kb/s. We use the EXG2 signal as the ground truth ECG in our experimental evaluation. We follow the same routine as in previous works~\cite{Xuesong2018,HR-CNN,DeepPhys} and use 30 seconds (frames 306 to 2135) of each video.


\textbf{Highly Compressed Videos.}\quad  Video compression was performed using the latest version of FFmpeg~\cite{ffmpeg}. We used three codecs (MPEG4, x264 and x265) in order to implement the three mainstream compression standards (H.263, H.264 and H.265). In order to demonstrate the effect of STVEN on highly compressed videos (i.e., with small file size and bitrates below 1000 kb/s), we compressed OBF videos into three qualities levels of average bitrate (file size) = 1000 kb/s (36.4 MB),500 kb/s (18.2 MB) and 250 kb/s (9.1 MB). The bitrates (file size) are about 20, 40 and 80 times smaller than those of original videos respectively.


\subsection{Implementation Details}
\vspace{-0.2em}

\textbf{Training Setting.}\quad For all facial videos, we use the Viola-Jones face detector ~\cite{viola2001rapid}  to detect and crop the coarse face area (see Figure~\ref{fig:featuremaps1} (a)) and remove background. We generate binary skin masks by open source Bob\footnote {https://gitlab.idiap.ch/bob/bob.ip.skincolorfilter} with threshold=0.3 as the ground truth. All face and skin images are normalized to 128x128 and 64x64 respectively.



The proposed method is trained in Nvidia P100 using PyTorch. The length of each video clip is $T=64$ while videos and ECG signals downsample into 30 fps and 30 Hz respectively. The partition for rPPGNet is $N=4$. The weights for different losses are set as $\alpha =1,\beta =0.5,\gamma =0.1,\delta =0.5$. As a part of the input, the compression bitrate label $k$ is represented by an one-hot mask vector. When joint training STVEN with rPPGNet, the loss balance weights $\varepsilon =1,\rho  =1e-4$. Adam optimizer is used while learning rate is set to 1$e-$4. We train rPPGNet for 15 epochs and STVEN for 20000 iterations. For the joint training, we fine-tuning STVEN for extra 10 epochs.


\textbf{Performance Metrics.}\quad  
For evaluating the accuracy of recovered rPPG signals, we follow previous works~\cite{li2018obf,Xuesong2018} and report both the average HR and several common HRV features on OBF dataset, and then evaluated several metrics of the average HR measurement on MAHNOB-HCI dataset. Four commonly used HRV features ~\cite{li2018obf,Poh2011} are calculated for evaluation, including respiratory frequency (RF) (in Hz), low frequency (LF), high frequency (HF) and LF/HF (in normalized units, n.u.). Both the recovered rPPGs and their corresponding ground truth ECGs go through the same process of filtering, normalization, and peak detection to obtain the inter-beat-intervals, from which the average HR and HRV features are calculated.

We report the most commonly used metrics for evaluating the performance, which include: the standard deviation (SD), the root mean square error (RMSE), the Pearson correlation coefficient (R), and the mean absolute error (MAE). $\bigtriangleup PSNR$ is also employed to evaluate changes of video quality before and after enhancement.


\subsection{Results on OBF}

OBF has large number of high quality video clips, which is suitable for verifying the robustness of our method in both average HR and HRV levels. We perform subject-independent 10-fold cross validation protocol to evaluate the rPPGNet and STVEN on the OBF dataset. At the testing stage, average HR and HRV features are calculated from output rPPG signals of 30 seconds length.

\begin{figure}
\includegraphics[width=8.8cm,height=8.5cm]{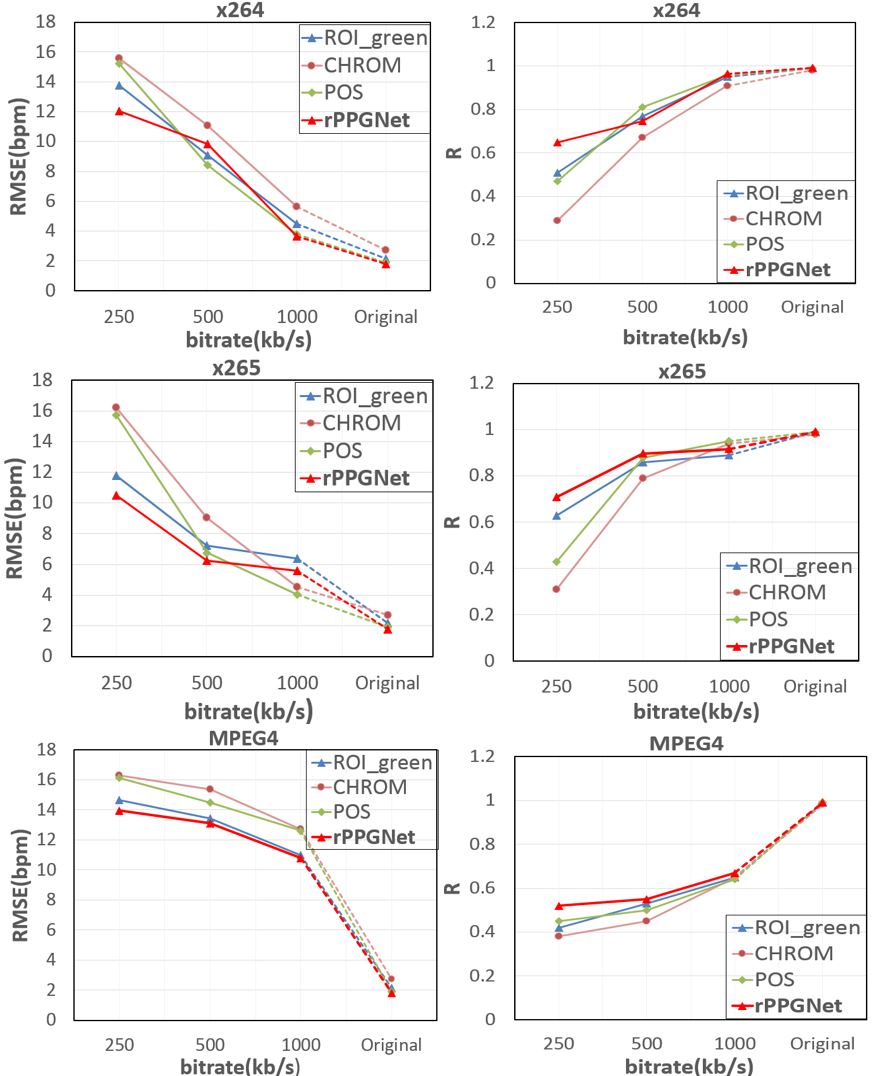}
  \caption{\small{HR measurement on OBF videos at different bitrates: all methods' performance drops with bitrates, while for the same bitrate level, the rPPGNet outperforms other methods.}} 
\label{fig:fig5}
\vspace{-1em}
\end{figure}


\textbf{Evaluation of rPPGNet on High Quality Videos.}\quad  Here, we re-implement several traditional methods~\cite{CHROM,li2018obf,wang2017algorithmic} on original OBF videos and compare the results in Table.~\ref{tab:OBFrPPGNet}. The results show  that rPPGNet (full) outperforms other methods for both averaged HR and HRV features. From ablation test results we can conclude that: 1) the skins segmentation module (the fifth row in Table.~\ref{tab:OBFrPPGNet}) slightly improves the performance with multi-task learning, which indicates these two tasks may have mutual hidden information. 2) The partition module (sixth row in Table.~\ref{tab:OBFrPPGNet}) further improves the performance by helping the model to learn more concentrated features. 3) Skin-based attention teaches the networks where to look and thus improves performance. In our observation, spatial attention with spatial-wise softmax operation works better than spatio-temporal attention, because in the rPPG recovery task the weights for different frames should be very close.





\textbf{Evaluation of rPPGNet on Highly Compressed Videos.}\quad   
We compressed OBF videos into three bitrates levels (250, 500 and 1000 kb/s) with three codecs (MPEG4, x264 and x265) as described in Section~\ref{sec:dataset}, so that we have nine groups (3 by 3) of highly compressed videos. We evaluate the rPPGNet together with three other methods on each of the nine groups of videos, using 10-folds cross-validation as before. 
The results are illustrated in Fig.~\ref{fig:fig5}. From the figure we can see that, first, the performance of both traditional methods and rPPGNet drop when bitrate decreases, which is true for all three compression codecs. The observation is consistent with previous findings\cite{mcduff2017impact,spetlik2018non} and proved that compression does impact rPPG measurement. 
Second, the important result is that when we compare at the same compression condition, rPPGNet can outperform other methods in most cases, especially very low bitrate of 250kb/s. This demonstrate the robustness of rPPGNet. 
But the accuracy at low bitrates is not satisfactory, and we hope to further improve the performance by video enhancement, i.e., using the proposed STVEN network.

\begin{figure}
 \vspace{-0.05em}
\centering

\includegraphics[width=8.5cm,height=3.2cm]{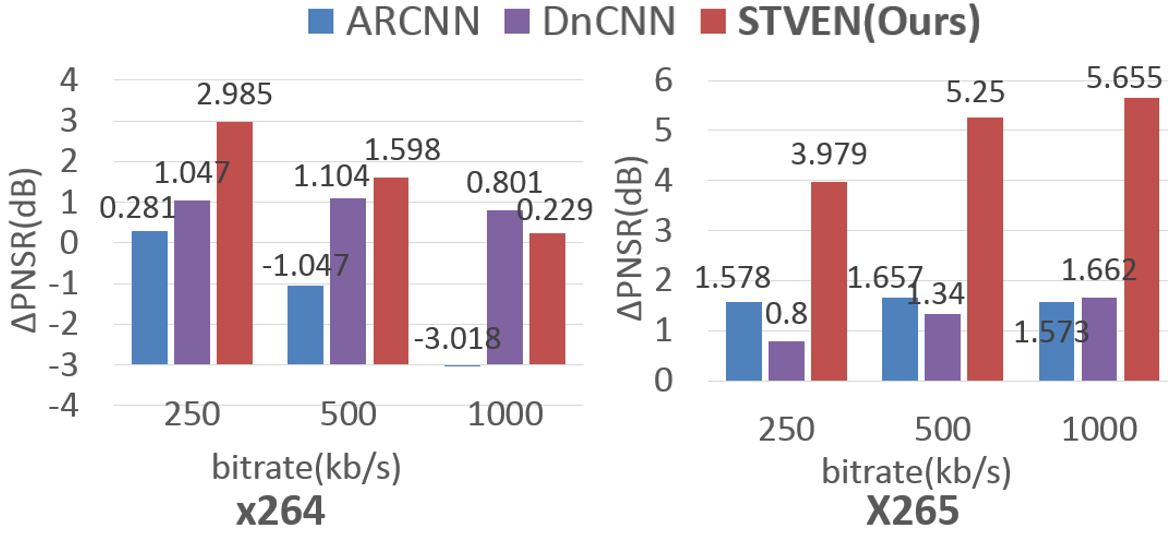}
  \caption{\small{Performance of video quality enhancement networks.}}
\label{fig:enhancement}
 \vspace{-0.5em}
\end{figure}

\begin{figure}
 \vspace{-0.05em}
\centering
\includegraphics[width=8.2cm,height=5.5cm]{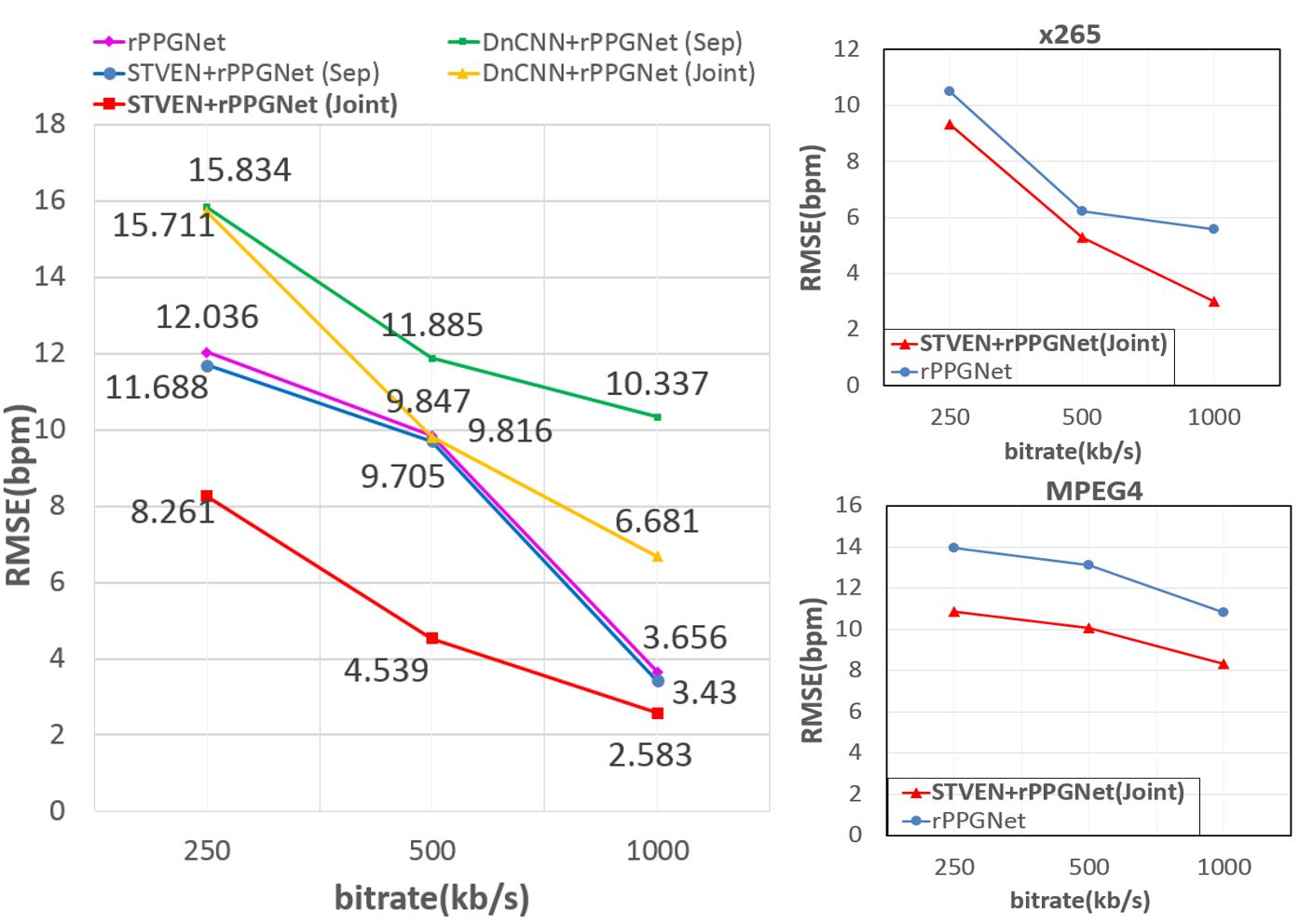}
  \caption{\small{HR measurement using different enhancement methods on highly compressed videos of OBF, left: with x264 codec; right: with x265 and MPEG4 codecs (cross-testing). Smaller RMSE indicates better performance}}
\label{fig:fig6}
 \vspace{-0.5em}
\end{figure}




\textbf{Evaluation of rPPGNet with STVEN for Enhancement on Highly Compressed Videos.}\quad Firstly, we demonstrate the STVEN does enhance the video quality on general level in terms of $\bigtriangleup PSNR$. As shown in Fig.~\ref{fig:enhancement}, the $\bigtriangleup PSNR$ of videos enhanced by STVEN are larger than zero, which indicate quality improvement. We also compared the STVEN to two other enhancement networks (ARCNN\cite{dong2015compression} and DnCNN\cite{zhang2017beyond}) and STVEN achieved even larger $\bigtriangleup PSNR$ than the other two methods.

Then we cascade STVEN with rPPGNet for verifying that the video enhancement model can boost performance of rPPGNet for HR measurement. We compare the performance of two enhancement networks (STVEN vs. DnCNN\cite{zhang2017beyond}) with two training strategies (separate training vs. joint training) on x264 compressed videos. Separate training means that the enhancement networks are pre-trained on highly compressed videos and the rPPGNet was pre-trained on high quality original videos, while joint training fine tunes the results of the two separate training with joint loss of the two tasks. 
The results in Fig.~\ref{fig:fig6}(left) shows that: for rPPG recovery and HR measurement on highly compressed videos, 1) STVEN helps to boost the performance of rPPGNet while DnCNN does not; and 2)joint training works better than separate training. It is surprising that STVEN boosts rPPGNet while DnCNN\cite{zhang2017beyond} suppresses rPPGNet in both separate training and joint training modes, which may be caused by the excellent spatio-temporal structure with fine-grained learning in STVEN and the limitation of the single-frame model of DnCNN. The generalization ability of STVEN-rPPGNet is shown in Fig.~\ref{fig:fig6}(right), in which the joint system trained on x264 videos was cross-tested on MPEG4 and x265 videos. Due to the quality and rPPG information enhancement by STVEN, rPPGNet is able to measure more accurate HR from untrained videos with MPEG4 and x265 compression.


\begin{table}[t]\small
\vspace{-0.3em}
\centering
\caption{Results of average HR measurement on MAHNOB-HCI.} \label{tab:ResultsMAHNOB}
 \begin{tabular}{l c c c c} 
 \toprule
 Method & HR$_{SD}$ & HR$_{MAE}$ & HR$_{RMSE}$ & HR$_{R}$\\
 & (bpm) & (bpm) & (bpm) & \\
 \midrule
 Poh2011~\cite{Poh2011} & 13.5 & - & 13.6 & 0.36 \\ 
 CHROM~\cite{CHROM} & - & 13.49 & 22.36 & 0.21 \\
 Li2014~\cite{Li2014} & 6.88 & - & 7.62 & 0.81\\
 SAMC~\cite{Tulyakov2016} & 5.81 & 4.96 & 6.23 & 0.83\\
 SynRhythm~\cite{Xuesong2018} & 10.88 & - & 11.08 & - \\ 
 HR-CNN~\cite{HR-CNN} & - & 7.25 & 9.24 & 0.51 \\
 DeepPhys~\cite{DeepPhys} & - & 4.57 & - & -\\
 
  \midrule
  rPPGNet & 7.82 & 5.51 & 7.82 & 0.78 \\
 STVEN+rPPGNet & \textbf{5.57} & \textbf{4.03} & \textbf{5.93} & \textbf{0.88}\\
 \bottomrule
 \end{tabular}
 \vspace{-1.5em}
\end{table}

 \vspace{-0.4em}
\subsection{Results on MAHNOB-HCI}
 \vspace{-0.3em}
 In order to verify the generalization of our method, we evaluate our methods on the MAHNOB-HCI dataset. MAHNOB-HCI is the most widely used dataset in HR measurement and the video samples are challenging because of the high compression rate and spontaneous motions, e.g., facial expressions. Subject-independent 9-fold cross validation protocol (3 subjects in a fold, totally 27 subjects) is adopted. As there are no original high quality videos available, the STVEN is trained with x264 highly compressed videos on OBF firstly and then cascades with the rPPGNet trained on MAHNOB-HCI for testing. Compared to the state-of-the-art methods in Table.~\ref{tab:ResultsMAHNOB}, our rPPGNet outperforms the deep learning based methods \cite{Xuesong2018, HR-CNN} in subject-independent protocol. With the help of video enhancement with richer rPPG information via STVEN, our two-stage method (STVEN+rPPGNet) surpasses all other methods. It indicates that STVEN can cross-boost the performance even when high-quality videos ground truth are not available.

 \vspace{-0.25em}
\subsection{Visualization and Discussion.}
 \vspace{-0.25em}
 
\begin{figure}
 \vspace{-0.05em}
\centering
\includegraphics[width=6.2cm,height=2.2cm]{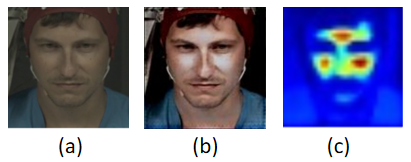}
  \caption{\small{Visualization of model output images. (a) face image in compressed video; (b) STVEN enhanced face image; (c) rPPGNet predicted attention map.}}
\label{fig:featuremaps1}
 \vspace{-1em}
\end{figure}

In Fig.~\ref{fig:featuremaps1}, we visualize an example to show the interpretability of our STVEN+rPPGNet method. The predicted attention map from rPPGNet Fig.~\ref{fig:featuremaps1}(c) focuses on the skin regions with strongest rPPG information (e.g., forehead and cheeks), which is in accordance with the priori knowledge mentioned in ~\cite{Verkruysse:08}. As shown in Fig.~\ref{fig:featuremaps1}(b), the STVEN enhanced face image seems to have richer rPPG information and stronger pulsatile flows in similar skin regions, which indicates the consistency of Fig.~\ref{fig:featuremaps1}(c).

\begin{figure}
\includegraphics[width=8.5cm,height=5cm]{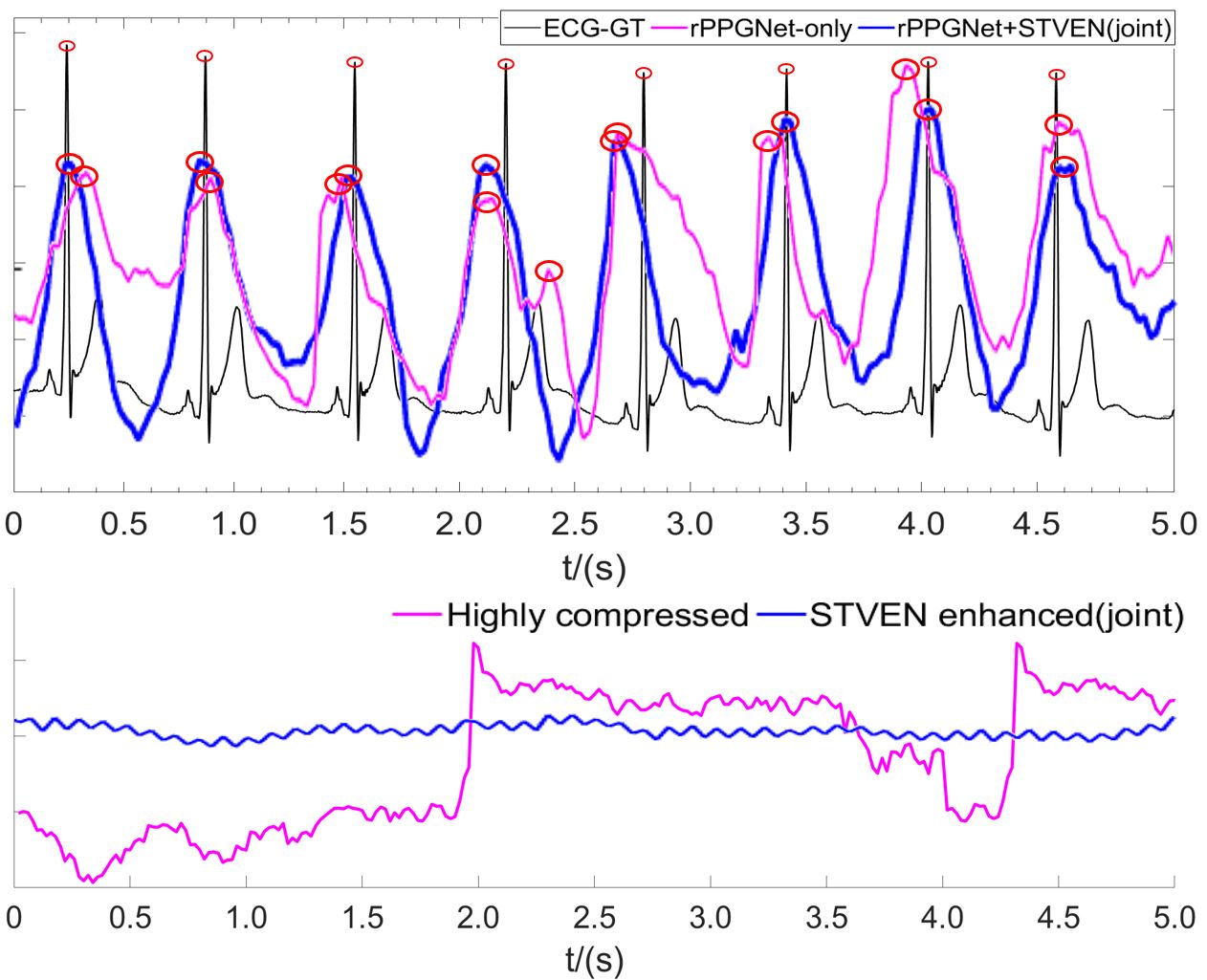}
  \caption{\small{ Predicted rPPG signals (top) and corresponding video PSNR curves (bottom). }}
\label{fig:qualityPSNR}
 \vspace{-0.5em}
\end{figure}

We also plot the rPPGNet recovered rPPG signals on highly compressed videos with and without STVEN. As shown in Fig.~\ref{fig:qualityPSNR}(top), benefited from the enhancement from STVEN, the predicted signals are with more accurate IBIs. Besides, Fig.~\ref{fig:qualityPSNR}(bottom) shows less objective quality (PSNR) fluctuation of the highly compressed videos with STVEN enhancement, which seems to help recover smoother and robust rPPG signals.

\vspace{-0.5em}
\section{Conclusions and Future Work}
\vspace{-0.1em}
\label{sec:conc}

In this paper, we proposed an end-to-end deep learning based method for rPPG signals recovery from highly compressed videos. In our method, the STVEN is firstly used to enhance the videos, and then the rPPGNet is cascaded to recover rPPG signals for HR and HRV features measurement. Comprehensive experiments are performed on two benchmark datasets and verified the effectiveness of the proposed method.
In future, we will try using compression related metrics like PSNR-HVS-M~\cite{ponomarenko2007between} to constrain the enhancement model STVEN. Moreover, we will also explore ways of building a novel metric for evaluating the video quality specially for the purpose of rPPG recovery.



\vspace{1em}
\noindent\textbf{Aknowledgement}\quad            This work was supported by the National Natural Science Foundation of China (No. 61772419), Tekes Fidipro Program
(No. 1849/31/2015), Business Finland Project (No. 3116/31/2017), Academy of Finland, and Infotech Oulu. As well, the authors wish to acknowledge CSC-IT Center for Science, Finland, for computational resources.


{\small
\bibliographystyle{ieee}
\bibliography{afinal}
}

\end{document}